%% file: main.tex
\documentclass[runningheads]{llncs}
\usepackage{graphicx}
\usepackage{subfigure} 
\usepackage{multirow} 
\usepackage{multicol} 
\usepackage{amsmath}
\usepackage{amsfonts}
\usepackage{bm}
\usepackage{adjustbox}
\usepackage{algorithm} 
\usepackage{algpseudocode} 
\usepackage[numbers,sort&compress]{natbib}
\usepackage{booktabs}
\usepackage{tabularx,pbox}
\usepackage{xcolor}
\usepackage{threeparttable}

\begin{document}
%
\title{Identifying Key Terms in Prompts for Relevance Evaluation with GPT Models}
%
%
\author{Jaekeol Choi}
\institute{{ Division of AI Data Convergence, Hankuk University of Foreign Studies, Seoul, South Korea\\}
jaekeol.choi@hufs.ac.kr }
%
\maketitle              

\input{0_Abstract}
\input{1_Introduction}
\input{2_Relatedworks}

\input{3_Methodology}
\input{4_Experiment}

\input{5_Discussion}

\input{6_Conclusion}

\section*{Acknowledgment}
This work was supported by Hankuk University of Foreign Studies Research Fund of 2024.

\appendix
\input{7_Appendix}

%
%
\bibliographystyle{splncs04nat}
\bibliography{bib}

\end{document}

%% file: 0_Abstract.tex
\begin{abstract}
Relevance evaluation of a query and a passage is essential in Information Retrieval (IR). Recently, numerous studies have been conducted on tasks related to relevance judgment using Large Language Models (LLMs) such as GPT-4, demonstrating significant improvements. However, the efficacy of LLMs is considerably influenced by the design of the prompt. The purpose of this paper is to identify which specific terms in prompts positively or negatively impact relevance evaluation with LLMs.
We employed two types of prompts: those used in previous research and generated automatically by LLMs. By comparing the performance of these prompts in both few-shot and zero-shot settings, we analyze the influence of specific terms in the prompts. 
We have observed two main findings from our study. 
First, we discovered that prompts using the term `answer' lead to more effective relevance evaluations than those using `relevant.' 
This indicates that a more direct approach, focusing on answering the query, tends to enhance performance. 
Second, we noted the importance of appropriately balancing the scope of `relevance.' 
While the term `relevant' can extend the scope too broadly, resulting in less precise evaluations, an optimal balance in defining relevance is crucial for accurate assessments. 
The inclusion of few-shot examples helps in more precisely defining this balance. By providing clearer contexts for the term `relevance,' few-shot examples contribute to refine relevance criteria.
In conclusion, our study highlights the significance of carefully selecting terms in prompts for relevance evaluation with LLMs.
\end{abstract}



\keywords{chatGPT, GPT-3.5, GPT-4, Information Retrieval, Large Language Models~(LLMs), relevance evaluation, prompt engineering,  passage ranking.}



%% file: 1_Introduction.tex
\section{Introduction}
\label{sec:intro}
Ranking models are foundational in the domain of Information Retrieval~(IR). Their success relies heavily on relevant sets that are used as standards during both training and testing stages.
Traditionally, crowd-sourced human assessors have been used for relevance judgement, as indicated by several studies~\citep{alonso2009can, blanco2011repeatable}. However, this method is often time-consuming, expensive, and can yield inconsistent results due to the inherent subjectivity of human judgement~\citep{maddalena2016crowdsourcing,nouri2020mining}. 

As technology keeps advancing, diverse machine learning techniques have stepped into the realm of relevance judgment~\citep{soboroff2001ranking,alonso2009can,carterette2006minimal,dietz2022wikimarks}. Driven by sophisticated algorithms, these methods tried to replicate or even enhance the human ability to discern relevance within vast information collections. 
Despite their potential, there remains skepticism among researchers about whether these techniques can match human accuracy and reliability in relevance judgment.

The major change came about with the advent of LLMs, notably GPT-3 and GPT-4. With their large architectures and extensive training datasets, these LLMs brought the possibility of automated relevance judgments. The performance of these models across diverse natural language processing tasks has fostered a renewed belief in the ability of machines to evaluate passage relevance accurately.
Encouraged by this paradigm shift, a couple of relevance judgment~\citep{ding2022gpt,faggioli2023perspectives} and ranking models~\citep{sun2023chatgpt} rooted in GPT architectures have been proposed. 
These models have demonstrated exceptional performance, often equaling or surpassing traditional methods.

However, The accuracy and robustness of relevance assessment using LLMs are significantly influenced by the prompts employed during the evaluation~\citep{lu2021fantastically,thomas2023large}. 
These prompts serve as critical guides, aligning the model's responses with the user's intent. 
Consequently, prompt formulation becomes a pivotal component, demanding careful design and optimization.

In this paper, we primarily focus on the prompts used for relevance evaluation in GPT models, particularly examining which terms in the prompts are beneficial or detrimental to performance. We investigate how the performance of LLMs varies with the use of different types of prompts: those utilized in previous research and those generated by LLMs. Our aim is to identify which terms in the prompts improve or impair the performance in relevance assessment tasks. To provide a comprehensive understanding, we conduct these experiments in both few-shot and zero-shot settings.

This study concludes that the term `answer' in prompt design is notably more effective than `relevant' for relevance evaluation tasks using LLMs. This finding emphasizes the importance of a well-calibrated approach to defining relevance. While `relevant' broadly encompasses various aspects of the query-passage relationship, `answer' more directly targets the core of the query, leading to more precise and effective evaluations. Therefore, balancing the scope of `relevance' in prompt design is crucial for enhancing the efficiency and accuracy of LLMs in relevance assessment.

The rest of this paper is organized as follows: `2 Related Work' delves into the background and previous studies. `3 Methodology' outlines the methods and approaches used in our study, including the details of the LLMs and the dataset. `4 Experimental Results' presents the findings from our experiments, providing a comprehensive analysis of the performance of different prompts. `5 Discussion' explore the implications of our findings. Finally, `6 Conclusions' summarizes the key insights from our study.

%% file: 2_Relatedworks.tex
\section{Related Work}

The field of IR has seen a significant evolution with the advent of advanced machine learning models and techniques. This section reviews the relevant literature, focusing on the development of relevance judgment methods in IR and the role of prompt engineering in the effective utilization of LLMs.

\subsection{Relevance Judgement in Information Retrieval}
The relevance evaluation between a query and a passage has been a fundamental task since the inception of ranking systems. This assessment has historically been conducted in a binary manner, categorizing results as either relevant or non-relevant, but has evolved to include graded relevance scales offering more detailed evaluations.

In the realm of traditional IR, the reliance on human assessors for relevance judgment has been extensively documented \citep{alonso2009can, blanco2011repeatable}. Despite their ability to provide nuanced evaluations, this approach has been criticized for its time and cost inefficiencies, as well as the subjective variability in results it can produce \citep{maddalena2016crowdsourcing,nouri2020mining}.

The advancement of machine learning and its integration into IR has marked a transition towards automated relevance judgment. This area, particularly the use of transformer-based models like BERT, has been the focus of recent research \citep{dietz2022wikimarks}. 
The challenge, however, lies in achieving a balance between the precision offered by human assessment and the scalability of automated methods.

The introduction of LLMs, especially GPT-3 and GPT-4, has further transformed the landscape of relevance judgment. Initial studies, such as those by \cite{wang2021want} and \cite{ding2022gpt}, explored the use of GPT-3 in annotation tasks, including relevance judgment. \citet{sun2023chatgpt}'s research extends this to examining GPT-3's broader capabilities in data annotation. In a distinct approach, \citet{macavaney2023one} investigated the use of LLMs for evaluating unassessed documents, aiming to improve the consistency and trustworthiness of these evaluations. Complementing this, \citet{thomas2023large} delved into the integration of LLMs for comprehensive relevance tagging, highlighting their comparable precision to human annotators. On the contrary, \citet{faggioli2023perspectives} has presented theoretical concerns regarding the exclusive use of GPT models for independent relevance judgment.

While extensive research has been conducted in this field, the specific influence of terms within a prompt on relevance evaluation remains unexplored. This study seeks to bridge this gap by investigating the impact of individual terms used in prompts.

\subsection{Few-shot and Zero-shot Approaches}
Recent advancements in LLMs have emphasized their capability for in-context learning, classified as either few-shot or zero-shot based on the presence of in-context examples. Few-shot learning, where a model is given a limited set of examples, has historically shown superior performance over zero-shot learning, which relies on instructions without examples, as highlighted by \citet{brown2020language}.

The ``pre-train and prompt'' paradigm emphasizes the distinction between few-shot prompts (conditioned on task examples) and zero-shot prompts (template-only). While few-shot learning was traditionally favored, recent studies, including those on GPT-4, suggest that zero-shot approaches can sometimes outperform few-shot methods, particularly in specific domains \cite{kojima2022large, openai2023gpt4}.

In our study, to investigate the terms in prompts, we conduct experiments using both few-shot and zero-shot settings and compare their outcomes.

\subsection{Advanced in Prompt Engineering}

Prompt engineering has emerged as a critical factor in harnessing the full potential of LLMs across various natural language processing applications. The formulation of a prompt is instrumental in guiding an LLM’s output, significantly influencing its performance in diverse tasks~\citep{schick2021few, brown2020language}. The art of crafting effective prompts involves meticulous design and strategic engineering, ensuring that prompts are precise and contextually relevant~\citep{reynolds2021prompt, gao2020making, shin2020autoprompt}.

The increasing complexity of LLMs has spurred interest in developing sophisticated prompt tuning methods. These methods often utilize gradient-based approaches to optimize prompts over a continuous space, aiming for maximal efficiency and efficacy~\citep{liu2023gpt, qin2021learning}. However, the practical application of these methods can be limited due to constraints such as restricted access to the models' gradients, particularly when using API-based models. This challenge has led to the exploration of discrete prompt search techniques, including prompt generation~\citep{ben2021pada}, scoring~\cite{yuan2021bartscore}, and paraphrasing~\cite{jiang2020can}.

In the broader context of prompt-learning, or ``prompting," the approach is increasingly recognized as a frontier in natural language processing, seamlessly bridging the gap between the pre-training and fine-tuning phases of model development~\citep{lester2021power, ding2021openprompt}. This technique is particularly valuable in low-data environments, where conventional training methods may be less effective~\citep{scao2021many, li2021sentiprompt, qin2021lfpt5}.

Within the realm of prompt-learning, two primary strategies are employed: few-shot and zero-shot learning. \citet{liang2022holistic} demonstrated a few-shot technique for generating relevance, while studies like those by \citet{sun2023chatgpt} and \citet{dai2022promptagator} have successfully applied few-shot learning in various scenarios. Conversely, \citet{ding2021openprompt} suggested that with an appropriate template, zero-shot prompt-learning could yield results surpassing those of extensive fine-tuning, emphasizing the power and flexibility of well-engineered prompts.

So far, there has been little focus on the terms within a prompt in existing research. This study is important because even small changes in a prompt can lead to different results. Our research, which concentrates on word terms, can be considered a form of micro-level prompt engineering.

%% file: 3_Methodology.tex
\section{Methodology}
\mathchardef\mhyphen="2D 
\begin{figure}[t]
\centering
\includegraphics[width=0.85\textwidth]{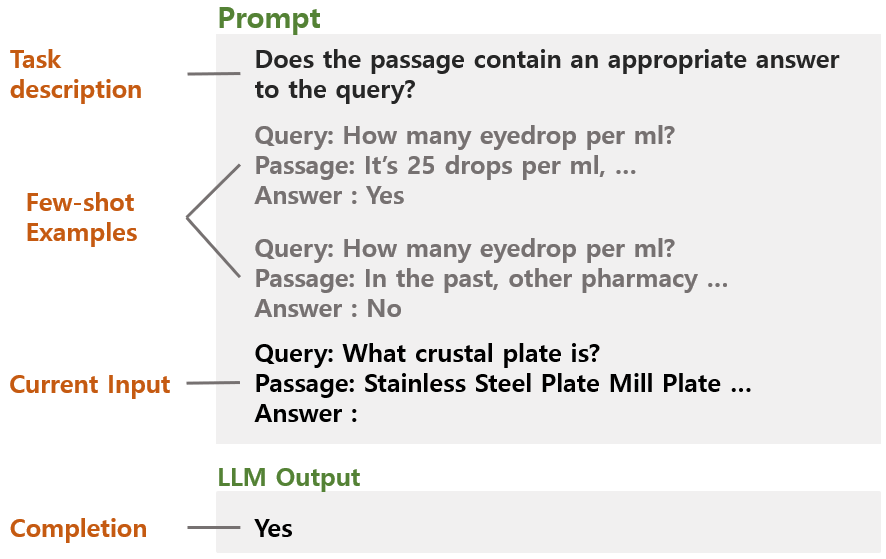}
\vspace*{8pt}
\caption{A prompt example for relevance evaluation. This example utilizes 2-shot examples.   \label{fig:one}}
\end{figure}

Prompts for relevance evaluation, as shown in Figure~\ref{fig:one}, include an instruction to guide the LLM, in-context few-shot examples for clarity, and an input as the target task. Using these elements, LLMs generate the corresponding output.
We apply this template in conduting our experiments for finding out which terms in prompts have impact on the performacne. 

\subsection{Evaluation method}
To evaluate the effectiveness of each prompt in the relevance evaluation task, an objective metric is required. 
For this purpose, we decided to use the similarity between the evaluations conducted by humans and those conducted by the LLM using the prompt. 
To measure the similarity between the two sets of evaluations, we utilize Cohen's kappa~($\kappa$) coefficient, a statistical measure for inter-rater reliability that accounts for chance agreement. 
This measure compares the agreement between relevance labels generated by the LLM and human judgments, reflecting the quality of the prompt. Higher kappa values indicate a stronger alignment between the LLM and human evaluations.
The Cohen's kappa coefficient is calculated using the following formula:
\begin{equation}
\kappa = \frac{P_o - P_e}{1 - P_e}
\end{equation}
In this equation, $P_o$ represents the observed agreement between the two sets of evaluations, and $P_e$ is the expected agreement by chance. The kappa value ranges from -1 to 1, where 1 indicates perfect agreement, 0 no agreement other than what would be expected by chance, and -1 indicates total disagreement. A higher kappa value suggests that the LLM's relevance evaluations are more closely aligned with human assessments, indicating a higher quality of the prompt in guiding the LLM to make evaluations similar to those of human judges.

\begin{table}[t]
\caption{Templates used for generating and analyzing by LLMs.\label{tab:prompt_template}}
\begin{tabularx}{\textwidth}{@{}ll@{}} 
\toprule
Usage & Template for generating prompts \\
\midrule
\multirow{5}{*}{Generation \;\;\;} & \pbox{0.8\textwidth}{Instruction: When given a query, a passage, and a few examples, generate a prompt that can make an output from the given input.}\\[0.3cm]
& Example 1 - {Input}:[query,passage]\verb|\n|{Output}:[Yes/No] \\
& Example 2 - {Input}:[query,passage]\verb|\n|{Output}:[Yes/No] \\
& ... \\[0.2cm]
& Generate prompt: \\
\midrule
\multirow{6}{*}{Analysis} & \pbox{0.8\textwidth}{Instruction: Which terms are common in these prompts that have a key role to evaluate relevance?} \\[0.3cm]
& Prompt 1: [Prompt] \\
& Prompt 2: [Promtp] \\
& ... \\[0.2cm]
& Find terms : \\
\bottomrule
\end{tabularx} 
\end{table}

\subsection{Prompts and Few-shot Examples}

We utilize two types of prompts, as shown in Table~\ref{tab:prompts} of Appendix B. The first type consists of prompts named with an `M', sourced from previous research~\citep{liang2022holistic,sun2023chatgpt,faggioli2023perspectives}. The second type includes prompts generated using the template in Table~\ref{tab:prompt_template}, which are named with a `G'. After assessing the performance of both prompt types, we aim to determine which prompts perform better. 
Following the experiments, we will analyze whether there are any terms common to the more effective prompts. 
If common terms are identified, it would suggest that these terms play a crucial role in the effectiveness of the prompt.

We conduct the experiments under both zero-shot and few-shot settings. Few-shot examples, derived from ~\citep{faggioli2023perspectives}, are illustrated in Table~\ref{tab:fewshots} of Appendix A. These few-shot examples consist of four instances: two are positive examples, and the other two are negative ones.
To ensure a fair comparison, we apply the same set of few-shot examples across all prompts.

\subsection{Analysis}

We analyze which terms are beneficial for relevance evaluation. Initially, we compare the performance of the prompts illustrated in Table~\ref{tab:prompts}. 
We then categorize the prompts into those with high performance and those with lower performance and look for distinguishing characteristics in each group. To identify the specific terms that play a role, we utilize the analysis prompts provided in Table~\ref{tab:prompt_template}. 
Furthermore, we compare how the results of each group vary depending on the presence or absence of few-shot examples.

We advance our analysis by constructing confusion matrices for the prompts, allowing for a more in-depth evaluation of their impact. 
Through the examination of precision and recall values derived from these matrices, we gain insights into the roles played by different terms within the context of relevance evaluation.


%% file: 4_Experiment.tex
\section{Experimental Result}
We presents the results of our experimental investigation into the effectiveness of various prompts in relevance evaluation tasks using LLMs. We detail the experimental setup, including the models and datasets used, and then delve into the outcomes of our experiments. These results provide crucial insights into how different prompt designs and key terms influence the performance of LLMs in relevance judgment tasks.

\subsection{Experimental Setup}
\subsubsection{Large Language Models}
For our experiments, we utilize GPT-3.5-turbo and GPT-4, both accessed via OpenAI's APIs. GPT-3.5-turbo, with its 178 billion parameters, enhances user interaction by providing clearer and more precise answers. As the most advanced in the series, GPT-4 has 1.76 trillion parameters and outperforms its predecessors in processing and contextual understanding.

\subsubsection{Dataset}
\begin{table}[t]
    \centering
    \caption{Overview of the TREC DL Passage datasets utilized in the study. The datasets from 2019 to 2021 are used for evaluating the performance of prompts. The table details the year of the dataset, the number of queries, the total number of query relevance judgments (qrels), and the number of sampled qrels used in the study.}
    \adjustbox{max width=\linewidth}{%
    \begin{tabular}{ccccc}
    \toprule
     Usage & TREC DL year & Number of  queries & Number of qrels & Number of sampeld qrels\\
    \midrule
    \multirow{3}{*}{Evaluation} & 2019  & 43 & 9,260 & 200 \\
    & 2020  & 54 & 11,386 & 200 \\
    & 2021  & 53 & 10,828 & 200 \\
    \bottomrule 
    \end{tabular}
    \label{tab:dataset} }
\end{table}
For our experiments, we utilize the test sets from the MS MARCO TREC DL Passage datasets spanning three years\footnote{https://microsoft.github.io/msmarco/TREC-Deep-Learning-2019 \ https://microsoft.github.io/msmarco/TREC-Deep-Learning-2020 \ https://microsoft.github.io/msmarco/TREC-Deep-Learning-2021}. 
As depicted in Table~\ref{tab:dataset}, We randomly sampled 200 data points from each year's test dataset, ensuring every query in the full set is included. These sampled datasets are then used to evaluate the prompts.

Relevance in these dataset is rated on a 4-point scale: ``Perfectly relevant,'' ``Highly relevant,'' ``Related,'' and ``Irrelevant.'' 

For binary classification tasks, we simplify this 4-point relevance scale to a binary ``Yes'' or "No'' judgment. Specifically, the categories of ``Perfectly relevant'' and ``Highly relevant'' are consolidated into a ``Yes'' category to indicate relevance, while ``Related'' and ``Irrelevant'' is classified as ``No.''

\subsection{Relevance Evaluation Result of Prompts }

\begin{table}[t]
    \centering
    \caption{Comparative Results of Relevance Evaluation in Zero-shot and Few-shot Settings: This table presents the performance of various prompts under zero-shot and few-shot scenarios. The top five performing prompts are highlighted in bold, while the bottom five are underlined. We provide the respective average performances for these groups in both GPT-3.5-turbo and GPT-4 models. A `*' symbol denotes a significant difference at the 95\% confidence level.  \label{tab:result}}
    \adjustbox{max width=\linewidth}{%
    \begin{tabular}{cccccc}
    \toprule 
    
    \multirow{2}{*}{\textbf{Type}} & \multirow{2}{*}{\textbf{\:\: Name \:\:}} & \multicolumn{2}{c}{\textbf{Zero-shot}} & \multicolumn{2}{c}{\textbf{Few-shot}} \\
    & & \textbf{GPT-3.5-turbo} & \textbf{\;\;\;GPT-4\;\;\;} & \textbf{GPT-3.5-turbo} & \textbf{\;\;\;GPT-4\;\;\;} \\
    \midrule
    \multirow{4}{*}{\textbf{Manual}} & \textbf{M1} & \textbf{0.389}~\small{($\pm$0.115)} & \textbf{0.450}~\small{($\pm$0.090)} & \textbf{0.339}~\small{($\pm$0.059)} & \textbf{0.471}~\small{($\pm$0.041)} \\
    & M2 & 0.326~\small{($\pm$0.032)} & 0.426~\small{($\pm$0.061)} & \underline{0.274}~\small{($\pm$0.064)} & 0.437~\small{($\pm$0.046)} \\
    & {M3} & \underline{0.319}~\small{($\pm$0.033)} & \underline{0.396}~\small{($\pm$0.086)} & {0.330}~\small{($\pm$0.025)} & 0.460~\small{($\pm$0.046)} \\
    & \underline{M4} & \underline{0.204}~\small{($\pm$0.019)} & \underline{0.344}~\small{($\pm$0.073)} & \underline{0.310}~\small{($\pm$0.041)} & \underline{0.433}~\small{($\pm$0.028)}\\
    \midrule 
    \multirow{10}{*}{\textbf{Generated}} & \underline{G1} & \underline{0.301}~\small{($\pm$0.046)} &  \underline{0.209}~\small{($\pm$0.116)} & \underline{0.309}~\small{($\pm$0.052)} & \underline{0.408}~\small{($\pm$0.029)}\\
     & {G2} & {0.356}~\small{($\pm$0.064)} & \underline{0.384}~\small{($\pm$0.099)} & 0.315~\small{($\pm$0.033)} & \underline{0.425}~\small{($\pm$0.050)} \\
     & \underline{G3} & \underline{0.279}~\small{($\pm$0.044)} & \underline{0.424}~\small{($\pm$0.060)} & \underline{0.303}~\small{($\pm$0.026)} & \underline{0.427}~\small{($\pm$0.067)}\\
     & G4 & \underline{0.268}~\small{($\pm$0.053)} & {0.426}~\small{($\pm$0.082)} & {0.312}~\small{($\pm$0.017)} & \underline{0.432}~\small{($\pm$0.054)}\\
     & G5 & {0.342}~\small{($\pm$0.007)} & 0.429~\small{($\pm$0.101)} & \underline{0.257}~\small{($\pm$0.031)} & 0.461~\small{($\pm$0.071)} \\
     & G6 & 0.363~\small{($\pm$0.085)} & \textbf{0.462}~\small{($\pm$0.073)} & \textbf{0.333}~\small{($\pm$0.073)} & \textbf{0.472}~\small{($\pm$0.046)}\\
     & \textbf{G7} & \textbf{0.393}~\small{($\pm$0.074)} &  \textbf{0.450}~\small{($\pm$0.066)} & \textbf{0.379}~\small{($\pm$0.042)} & \textbf{0.464}~\small{($\pm$0.051)} \\
     & \textbf{G8} & \textbf{0.382}~\small{($\pm$0.075)} & \textbf{0.455}~\small{($\pm$0.084)} & \textbf{0.349}~\small{($\pm$0.066)}  & \textbf{0.463}~\small{($\pm$0.039)} \\
     & \textbf{G9} & \textbf{0.398}~\small{($\pm$0.089)} & \textbf{0.443}~\small{($\pm$0.074)} & \textbf{0.351}~\small{($\pm$0.078)} & \textbf{0.468}~\small{($\pm$0.046)} \\
     & {G10} & \textbf{0.366}~\small{($\pm$0.086)} & {0.442}~\small{($\pm$0.074)} & 0.327~\small{($\pm$0.050)} & 0.445~\small{($\pm$0.055)} \\
    \midrule
    \multicolumn{2}{c}{\textbf{Top-5 average}} & {0.386}~\small{($\pm$0.013)}$^*$ & {0.452}~\small{($\pm$0.007)}$^*$ & 0.352~\small{($\pm$0.018)}$^*$ & 0.468~\small{($\pm$0.004)}$^*$ \\
    \multicolumn{2}{c}{\textbf{Bottom-5 average}} & 0.274~\small{($\pm$0.044)} & 0.351~\small{($\pm$0.084)} & 0.291~\small{($\pm$0.024)} & 0.425~\small{($\pm$0.010)} \\
    \bottomrule
    \end{tabular} }
\end{table}

The evaluation of prompt efficacy in relevance assessments, as outlined in Table~\ref{tab:result}, reveals notable trends. A key observation is the significant performance variation among semantically similar prompts, highlighting the impact of subtle differences in prompt design on evaluation outcomes. For example, although M3 and G3 are similar prompts asking if the query and passage are `relevant,' they yield different results. Moreover, despite all prompts addressing the relevance between the query and passage, their outcomes vary substantially.

When comparing results between GPT-3.5 and GPT-4 across both few-shot and zero-shot settings,
Prompts M1, G7, G8, and G9 consistently rank in the top five across both GPT-3.5-turbo and GPT-4, indicating their inherent effectiveness. 
Conversely, certain prompts consistently underperform in both models. Specifically, prompts M4, G1, and G3 are found in the bottom five, underscoring elements that may detract from the efficacy of relevance evaluations.

Examining the performance of individual models reveals distinct characteristics in response to the prompts. 
Each model demonstrates unique preferences in prompt efficacy, illustrating that LLMs may respond differently to the same prompt structures. Certain prompts show high efficacy in GPT-3.5-turbo, while others perform better in GPT-4. 
Notably, GPT-4 generally exhibits superior performance compared to GPT-3.5-turbo across a range of prompts. 
A particular case of interest is prompt G1 in the zero-shot setting, where GPT-4's performance is the only instance of falling behind GPT-3.5-turbo.
Aside from this case, GPT-4's performance is generally superior to that of GPT-3.5-turbo.

Further statistical analysis, involving a paired t-test on the averages of the top five and bottom five prompts, reinforces these findings. Specifically, the top five prompts in GPT-3.5-turbo had an average performance of 0.386, while in GPT-4, this average was higher at 0.452. Conversely, the bottom five prompts averaged 0.274 in GPT-3.5-turbo and 0.351 in GPT-4. These results indicate a statistically significant difference in performance at a 95\% confidence level, emphasizing the pivotal role of prompt design in influencing the effectiveness of relevance evaluations in LLMs.

\subsection{Analysis of Terms in prompts}

\begin{table}[t]
    \centering
    \caption{Key terms that have an crucial role. In prompts demonstrating good performance, the term `answer' is commonly used, whereas in prompts indicating low performance, the term `relevant' is commonly used.}
    \begin{tabular}{ccl}
    \toprule
    \textbf{Efficacy} & \textbf{Key Term} & \textbf{Prompt} \\
    \midrule
    \multirow{4}{*}{High} & \multirow{4}{*}{\textbf{Answer}} & \pbox{0.7\textwidth}{G9: ... if the passage provides a direct \textbf{answer} to ... } \\
    & &  \pbox{0.7\textwidth}{G7: ... the passage contains the \textbf{answer} to the query ... }\\
    & & \pbox{0.7\textwidth}{M1: Does the passage \textbf{answer} the query? ... } \\
    & & \pbox{0.7\textwidth}{G10: Determine if the passage correctly \textbf{answers} to ... } \\
    \midrule
    \multirow{4}{*}{Low} & \multirow{4}{*}{\textbf{Relevant}} & \pbox{0.7\textwidth}{G1: Do the query and passage \textbf{relate} to the same topic.. } \\
    & & \pbox{0.7\textwidth}{M4: 2 = highly \textbf{relevant}, very helpful for ... } \\
    & & \pbox{0.7\textwidth}{M3: Indicate if the passage is \textbf{relevant} for the query? ...} \\
    & & \pbox{0.7\textwidth}{G3: In the context of the query, is the passage \textbf{relevant}? } \\ 
    \bottomrule
    \end{tabular}
    \label{tab:term} 
\end{table}

In our analysis, we utilized the template from Table~\ref{tab:prompt_template} to identify key terms in prompts that play a significant role in relevance evaluation using LLMs. The findings are summarized in Table~\ref{tab:term}.

We observed that prompts demonstrating top performance commonly used the term `answer' or its variations. For instance, in M1, the prompt asks if the passage `answers' the query. Similarly, G7 and G9 emphasize whether the passage contains or directly `answers' the query. This pattern is also evident in G10, where the prompt focuses on whether the passage `correctly answers' the query.

On the other hand, prompts associated with lower performance frequently included the term `relevant' or related terms. For example, M3's prompt requires indicating if the passage is `relevant' for the query, while G1 asks if the query and passage `relate' to the same topic. This trend continues in M4 and G3, where the term `relevant' is central to the prompt's structure.

These findings indicate that the choice of key terms in prompts significantly impacts the performance of LLMs in relevance evaluation tasks. Terms like `answer' seem to guide the LLM towards more effective evaluation, while the use of `relevant' appears to be less conducive for this purpose.

\subsection{Analysis of Zero-shot and Few-shot Results}

\begin{figure}[t]
\centering
\includegraphics[width=0.85\textwidth]{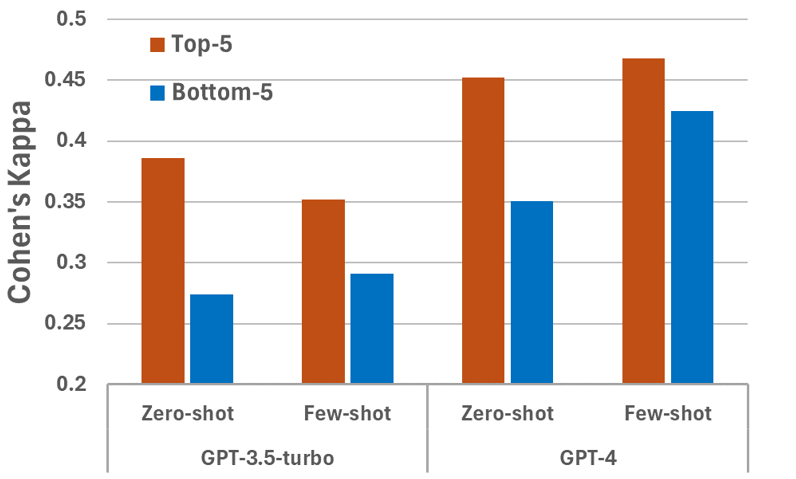}
\vspace*{8pt}
\caption{Average Cohen's kappa values for top-5 and bottom-5 prompts in GPT-3.5-turbo and GPT-4 across few-shot and zero-shot settings.   \label{fig:two}}
\end{figure}

The differences in performance between zero-shot and few-shot models for GPT-3.5-turbo and GPT-4 are illustrated in Figure~\ref{fig:two}, which presents the average results for each approach. From this analysis, we can discern two interesting observations.

Firstly, there is a notable variation in performance across the top and bottom five performers between the two model versions. In the case of GPT-3.5-turbo, while there is an improvement in the performance of the bottom five prompts (from an average of 0.274 in zero-shot to 0.291 in few-shot), the top five prompts exhibit a decrease in performance (from 0.386 in zero-shot to 0.352 in few-shot). 
This indicates that while few-shot examples enhance GPT's ability to handle previously lower-performing prompts, they might detrimentally affect the performance of the highest-performing prompts.

In contrast, GPT-4 shows a consistent improvement in both the top and bottom performers with few-shot examples. The top five prompts improve from an average of 0.452 in zero-shot to 0.468 in few-shot, and the bottom five improve from 0.351 to 0.425. 
This shows that few-shot examples enhance the overall performance in evaluation tasks with GPT-4.

Secondly, both models demonstrate a reduction in the performance gap between the top and bottom five prompts with few-shot learning. 
This convergence is more pronounced in GPT-4, which sees a more significant increase in performance for the bottom five prompts. 
It suggests that few-shot examples is particularly effective in refining the model's responses to less optimal prompts, leading to a more consistent performance across different types of prompts.

Given the role of few-shot examples in providing clearer instructions and context, these results suggest that GPT-4 is more adept at adapting to varied prompt structures and content than GPT-3.5-turbo.

%% file: 5_Discussion.tex
\section{Discussion}
This section offers an analysis of our experimental results, focusing on the impact of specific prompt terms on the performance of LLMs in relevance evaluation. We also discuss the potential and challenges of using LLMs as direct rankers in IR, compared to their current role in generating relevance judgments.

\begin{table}[t]
    \centering
    \caption{Confusion Matrices for three prompts using the TREC DL 2021 test set  in a zero-shot setting. This table includes Cohen's kappa values, along with calculated precision and recall. The analysis focuses on G6 with the highest performance, G1 with the lowest, and G10, which has the narrowest definition by using the term of `correctly'.}

\begin{tabular}{ccccccc}
\toprule
\multirow{2}{*}{Prompt} & \multirow{2}{*}{Prediction} & \multicolumn{2}{c}{Human assessors} & \multirow{2}{*}{Cohen's $\kappa$} & \multirow{2}{*}{Precision} & \multirow{2}{*}{Recall} \\
    &  & Relevant & Irrelevant & \\
\midrule
 \multirow{2}{*}{\textbf{G6}} &  Relevant & \bf{{43}} &  {24} & \multirow{2}{*}{\textbf{0.528}} & \multirow{2}{*}{{0.641}} & \multirow{2}{*}{{0.716}} \\
  & Irrelevant & 17 & \bf{116} & \\
\midrule
\multirow{2}{*}{\underline{G1}} &  Relevant & \bf{{59}} &  {84} &  \multirow{2}{*}{\underline{0.275}} & \multirow{2}{*}{\underline{0.413}} & \multirow{2}{*}{{\textbf{0.983}}}\\
 & Irrelevant & 1 & \bf{56} &  \\
\midrule
\multirow{2}{*}{{G10}} &  Relevant & \bf{{38}} &  {20} &  \multirow{2}{*}{0.495 } & \multirow{2}{*}{\textbf{0.655}} & \multirow{2}{*}{\underline{0.633}}\\
 & Irrelevant & 22 & \bf{120} & \\
\bottomrule
\end{tabular}
    \label{tab:gpt4_confusion}
\begin{flushleft}
\footnotesize{G6\;\;: Given a query and a passage, determine if the passage provides \textbf{an answer} to the query. ... }  \\
\footnotesize{G1\;\;: Do the query and passage \textbf{relate} to the same topic? ... } \\
\footnotesize{G10\;: Determine if the passage \textbf{correctly answers} a given query. ...} \\
\end{flushleft}
\end{table}

\subsection{Why `Answer' Is Better Than `Relevant'}
The analysis of confusion matrices in Table~\ref{tab:gpt4_confusion} provides key insights into the effectiveness of different prompt types in relevance evaluation. This analysis highlights G6, which had the highest performance, G1 with the lowest performance, and G10, known for its use of the term `correctly.'

G6, achieving the highest performance, questions if the passage provides `an answer' to the query. This prompt led to a significant agreement between LLM predictions and human assessors, as evident by a high Cohen's kappa value of 0.528, along with strong precision and recall. The high number of true positives (43) and true negatives (116) in G6's matrix suggests that focusing on `answering' is highly effective in evaluating the relevance of the passage to the query.

Conversely, G1, which demonstrated the lowest performance, focuses on whether the query and passage `relate' to the same topic. Despite its high recall, this prompt yielded a lower Cohen's kappa value of 0.275. The comparatively fewer true negatives (56) against G6 indicate that a broader `relevance' focus may lead to less precise evaluations.

G10, with its emphasis on whether the passage `correctly answers' the query, shows a distinct performance, marked by a Cohen's kappa value of 0.495. Its precision is notably high, but the recall is somewhat limited, suggesting that while it is effective in identifying specific relevant answers, it may overlook some broader aspects of relevance.

This comparison underlines the varying effectiveness of prompts based on their focus in the context of information retrieval. Prompts like G6, with an `answering' focus, tend to lead to more accurate and precise evaluations, while `relevance'-focused prompts like G1 might not capture the entire scope of the query-passage relationship. G10's specific focus on `correctly answering' demonstrates a particular effectiveness in identifying precise answers but at the potential expense of broader relevance. Therefore, the choice of key terms and their emphasis is crucial in designing prompts for efficient retrieval and ranking in LLMs.

\subsection{Balancing the Definition of `Relevance'}

As discussed in the previous section, defining `relevance' in the context of LLM prompts varies significantly in its scope. G10's approach, using the term `correctly answers', tends to give a slightly narrow definition in relevance evaluation. It focuses on whether the passage precisely addresses the query, potentially overlooking broader aspects of relevance.

On the other hand, we explored a more balanced approach with G6's prompt. This prompt, focusing on whether the passage provides `an answer' to the query, strikes a middle ground. It covers not just the direct answer but also the broader context, leading to a more comprehensive consideration of relevance.

Conversely, G1's prompt offers the broadest definition of relevance by asking if the query and passage `relate' to the same topic. This wide approach, while inclusive, risks being too expansive. As reflected in the confusion matrix for G1 in Table~\ref{tab:gpt4_confusion}, this broad definition results in high recall but at the cost of lower precision, as it captures a wide net of potentially relevant information, including false positives.

This analysis highlights the need for a balanced definition of relevance in prompt design. While G1's broad approach increases recall, its precision suffers. G10's narrow focus may miss broader relevance aspects. In contrast, G6’s approach appears to offer a more optimal balance. It captures a wide array of relevant information without being overly narrow or inclusive, leading to more accurate and balanced performance in relevance evaluations. These findings are pivotal for crafting prompts that precisely measure the relevance of information in LLM-based retrieval and ranking systems.

\subsection{Influence of Few-shot Examples}
As can be seen in Figure~\ref{fig:two}, in GPT-3.5-turbo, the performance of zero-shot is slightly higher than that of few-shot. In contrast, in GPT-4, the performance of few-shot exceeds that of zero-shot. 
This variation indicates that a conclusive determination of the relative impacts of few-shot and zero-shot approaches is complex and model-dependent.

However, there is a characteristic that appears consistently in both models: the use of few-shot examples reduces the performance gap between the top-5 and bottom-5 groups. In GPT-3.5-turbo, the gap decreased from 0.112 to 0.061, and in GPT-4, it nearly halved from 0.101 to 0.043. These results suggest that few-shot examples help in defining unclear aspects in the bottom-5 instructions.
For instance, consider the case of the G1 prompt. In the zero-shot setting, GPT-4 shows a low performance of 0.209, but when few-shot examples are used, the performance dramatically increases to 0.409. This could indicate that while the term `relate' in G1 has a broad meaning, the use of few-shot examples helps in clarifying its interpretation.


\subsection{Direct Ranking vs. Relevance Judgment Using LLMs}

An emerging area of interest is the potential for using LLMs directly as rankers in IR, rather than just for generating relevance judgments.
However, the practical application of LLMs as direct rankers faces significant challenges, primarily due to efficiency concerns. Directly ranking with LLMs, especially when reliant on API calls, can be slow and costly, as it requires repeated, resource-intensive interactions with the model for each ranking task. This approach, therefore, becomes impractical for large-scale or real-time ranking applications.

Given these constraints, future research in this domain should consider the development and utilization of downloadable, standalone LLMs. Such models, once sufficiently advanced, could potentially be integrated directly into ranking systems, offering a more efficient and cost-effective solution compared to API-dependent models. This shift would allow for the direct application of LLMs in ranking tasks, potentially overcoming the limitations currently posed by API reliance. However, this path also necessitates further advancements in LLM technology to ensure these models can operate effectively and reliably in a standalone capacity.





%% file: 6_Conclusion.tex
\section{Conclusions}

In this paper, we have examined the nuances of prompt design in relevance evaluation tasks using Large Language Models such as GPT-3.5-turbo and GPT-4. Our research reveals the profound impact that specific terms within prompts have on the effectiveness of these models. Contrary to initial expectations, our findings indicate that prompts focusing on `answering' the query are more effective than those emphasizing broader concepts of `relevance.' This highlights the importance of precision in relevance assessments, where a direct answer often more closely aligns with the intended query-passage relationship.

Furthermore, our investigations into few-shot and zero-shot scenarios revealed contrasting impacts on model performance. We found that few-shot examples tend to enhance the performance of LLMs, particularly in GPT-4, by bridging performance gaps between differently functioning prompts.

Our study also underscores the need for a well-balanced definition of `relevance' in prompt design. We observed that overly broad definitions, while helpful in increasing recall, can compromise precision. Conversely, narrowly defined prompts, though precise, risk missing broader relevance aspects, failing to capture a comprehensive relevance assessment. Therefore, striking the right balance in prompt design is crucial for enhancing the efficiency and accuracy of LLMs in relevance evaluation tasks.

In summary, this paper contributes to the field by providing new insights into optimizing LLMs for relevance evaluation tasks. These insights offer crucial guidelines for creating effective prompts, ensuring that LLM outputs align more accurately with nuanced, human-like relevance judgments. 
As LLM technology continues to evolve, understanding the subtleties of prompt design becomes increasingly important in natural language processing and information retrieval applications.

%% file: 7_Appendix.tex
\section{Few-shot Exmaples}
We utilize four few-shot exmaples for our experiments.

\begin{table}[t]
\caption{Four few-shot exsmples \label{tab:fewshots}}
\begin{tabularx}{\textwidth}{@{}cc@{}} 
\toprule
  \# & Few-shot examples \\
\midrule
1 \:\: & \pbox{0.93\textwidth}{Query: how many eye drops per ml \\
Passage: Its 25 drops per ml, you guys are all wrong. If it is water, the standard was changed 15 - 20 years ago to make 20 drops = 1mL. The viscosity of most things is temperature dependent, so this would be at room temperature. Hope this helps. \\
Answer: Yes \\
} \\
2 \:\: & \pbox{0.93\textwidth}{Query: how many eye drops per ml \\
Passage: RE: How many eyedrops are there in a 10 ml bottle of Cosopt? My Kaiser pharmacy insists that 2 bottles should last me 100 days but I run out way before that time when I am using 4 drops per day.In the past other pharmacies have given me 3 10-ml bottles for 100 days.E: How many eyedrops are there in a 10 ml bottle of Cosopt? My Kaiser pharmacy insists that 2 bottles should last me 100 days but I run out way before that time when I am using 4 drops per day. \\
Answer: No \\
} \\
 3 \:\: & \pbox{0.93\textwidth}{Query: can you open a wells fargo account online \\
Passage: You can transfer money to your checking account from other Wells Fargo. accounts through Wells Fargo Mobile Banking with the mobile app, online, at any. Wells Fargo ATM, or at a Wells Fargo branch. 1 Money in — deposits. \\
Answer: No \\
} \\
 4 \:\: & \pbox{0.93\textwidth}{Query: can you open a wells fargo account online \\
Passage: You can open a Wells Fargo banking account from your home or even online. It is really easy to do, provided you have all of the appropriate documentation. Wells Fargo has so many bank account options that you will be sure to find one that works for you. They offer free checking accounts with free online banking. \\
Answer: Yes \\
} \\
\bottomrule
\end{tabularx} 
\end{table}
\section{Prompts}
We utilize 14 prompts for our experiments. 

\begin{table}[t]
\caption{List of 14 prompts used in the experiments, detailing their names and instructions. \label{tab:prompts}}
\begin{tabularx}{\textwidth}{@{}cll@{}} 
\toprule
 \multicolumn{2}{c}{\;\; \textbf{Name} \;} & \textbf{Prompt instruction} \\
\midrule
 \multirow{11}{*}{\rotatebox{90}{\textbf{Manual}}} & M1 & \pbox{0.84\textwidth}{Does the passage answer the query? Respond with `Yes' or `No'.} \\[0.2cm] 
 & M2 & \pbox{0.84\textwidth}{Given a passage and a query, predict whether the passage includes an answer to the query by producing either ``Yes'' or ``No''. } \\[0.3cm] 
 & M3 & \pbox{0.84\textwidth}{Indicate if the passage is relevant for the query. Respond with ``Yes'' or ``No''.} \\[0.3cm]  
 & M4 & \pbox{0.84\textwidth}{You are a search quality rater evaluating the relevance of passages. Given a query and a passages, you must provide a score on an integer scale of 0 to 2 with the following meanings: \\ 2 = highly relevant, very helpful for this query \\ 1 = relevant, may be partly helpful but might contain other irrelevant content \\ 0 = not relevant, should never be shown for this query } \\
\midrule
\multirow{20}{*}{\rotatebox{90}{\textbf{Generated}}} & G1 & \pbox{0.84\textwidth}{Do the query and passage relate to the same topic? Respond with `Yes' or `No'.} \\[0.2cm]  
 & G2 & \pbox{0.84\textwidth}{Is the passage pertinent to the query? Indicate with `Yes' or `No'.} \\[0.3cm]
 & G3 & \pbox{0.84\textwidth}{In the context of the query, is the passage relevant? Reply with `Yes' or `No'.} \\[0.2cm]
 & G4 & \pbox{0.84\textwidth}{Would a user find the passage relevant to their query? Respond with `Yes' or `No'.} \\[0.2cm]
 & G5 & \pbox{0.88\textwidth}{Does the passage contain information relevant to the query? Answer with `Yes' or `No'.} \\[0.2cm]
 & G6 & \pbox{0.84\textwidth}{Given a query and a passage, determine if the passage provides an answer to the query. If the passage answers the query, respond with ``Yes''. If the passage does not answer the query, respond with ``No''.} \\[0.5cm]
 & G7 & \pbox{0.85\textwidth}{Your task is to determine whether the passage contains the answer to the query or not. If the passage contains the answer to the query, your response should be `Yes'. If the passage does not contain the answer, your response should be `No'} \\[0.5cm] 
 & G8 & \pbox{0.84\textwidth}{Given a query and a passage, determine if the passage provides a satisfactory answer to the query. Respond with `Yes' or `No'.} \\[0.4cm]
 & G9 & \pbox{0.84\textwidth}{Given a query and a passage, determine if the passage provides a direct answer to the query. Answer with `Yes' or `No'} \\[0.3cm]
 & G10 & \pbox{0.86\textwidth}{Determine if the passage correctly answers a given query. Respond with `Yes' or `No'} \\
\bottomrule
\end{tabularx} 
\end{table}